\documentclass[12pt,preprint]{aastex}
\usepackage{color}

\def\lbqs{LBQS 0103-2753}
\def\cosmos{COSMOS 100043.15+020637.2}

\def\etal{~et al.}
\def\simlt{\lower.5ex\hbox{$\; \buildrel < \over \sim \;$}}
\def\simgt{\lower.5ex\hbox{$\; \buildrel > \over \sim \;$}}

\def\gsim{\lower 2pt \hbox{$\, \buildrel {\scriptstyle >}\over
{\scriptstyle \sim}\,$}}
\def\lsim{\lower 2pt \hbox{$\, \buildrel {\scriptstyle <}\over
{\scriptstyle \sim}\,$}}

\def\deg{\ifmmode ^{\circ}
         \else $^{\circ}$\fi}
\def\pdeg{\ifmmode
           $\setbox0=\hbox{$^{\circ}$}\rlap{\hskip.11\wd0 .}$^{\circ}
     \else \setbox0=\hbox{$^{\circ}$}\rlap{\hskip.11\wd0 .}$^{\circ}$\fi}
     
\def\pc{\ifmmode \mathrm{pc} \else $\mathrm{pc}$ \fi}
\def\mpc{\ifmmode \mathrm{Mpc} \else $\mathrm{Mpc}$\fi}
\def\mpcthree{\ifmmode \mathrm{Mpc}^{-3} \else $\mathrm{Mpc}^{-3}$\fi}
\def\gpcthree{\ifmmode \mathrm{Gpc}^{-3} \else $\mathrm{Gpc}^{-3}$\fi}

\def\kelvin{\ifmmode \mathrm{K} \else {$\mathrm{K}$}\fi}
\def\kev{\ifmmode \mathrm{keV} \else $\mathrm{keV}$ \fi}

\def\lsun{\ifmmode {L_\odot} \else $L_\odot$\fi}
\def\msun{\ifmmode M_\odot \else $M_\odot$\fi}
\def\msunyr{\ifmmode M_\odot~\mathrm{yr}^{-1} \else $M_\odot~\mathrm{yr}^{-1}$\fi}

\def\cosi{\ifmmode {\cos\,i} \else $\cos\,i$\fi}

\def\heii{\ifmmode {\rm He{\sc ii}} \else He~{\sc ii}\fi}
\def\mgii{\ifmmode {\rm Mg{\sc ii}} \else Mg~{\sc ii}\fi}
\def\caii{\ifmmode {\rm Ca{\sc ii}} \else Ca~{\sc ii}\fi}
\def\ciii{\ifmmode {\rm C{\sc iii}]} \else C~{\sc iii}]\fi}
\def\civ{\ifmmode {\rm C{\sc iv}} \else C~{\sc iv}\fi}
\def\mgii{\ifmmode {\rm Mg{\sc ii}} \else Mg~{\sc ii}\fi}

\newcommand{\oiii}{{\sc [O~iii]}}
\newcommand{\ovi}{{\sc O~vi}}

\def\teff{\ifmmode {T_{\rm eff}} \else $T_{\rm eff}$\fi}
\def\tmax{\ifmmode {T_{\rm max}} \else $T_{\rm max}$\fi}

\def\mbh{\ifmmode {M_{\rm BH}} \else $M_{\rm BH}$\fi}
\def\led{\ifmmode L_{\mathrm{Ed}} \else $L_{\mathrm{Ed}}$\fi}
\def\lbolflare{\ifmmode L_{\mathrm{bol,flare}} \else $L_{\mathrm{bol,flare}}$\fi}
\def\lagn{\ifmmode L_{\mathrm{agn}} \else $L_{\mathrm{agn}}$\fi}
\def\lbolagn{\ifmmode L_{\mathrm{bol,agn}} \else $L_{\mathrm{bol,agn}}$\fi}
\def\lbol{\ifmmode L_{\mathrm{bol}} \else $L_{\mathrm{bol}}$\fi}
\def\mdot{\ifmmode {\dot M} \else $\dot M$\fi}
\def\mdoto{\ifmmode {\dot{M}_0} \else  $\dot{M}_0$\fi}
\def\mdotf{\ifmmode {\dot{M}_\mathrm{flare}} \else  $\dot{M}_\mathrm{flare}$\fi}

\def\hst{{\em HST}}
\def\hnot{\ifmmode H_0 \else H$_0$ \fi}

\def\vkep{\ifmmode v_\mathrm{Kep} \else $v_\mathrm{Kep}$ \fi}
\def\vc{\ifmmode v_\mathrm{c} \else $v_\mathrm{c}$ \fi}

\def\vthree{\ifmmode v_{1000} \else $v_{1000}$ \fi}
\def\vrel{\ifmmode v_\mathrm{rel} \else $v_\mathrm{rel}$ \fi}
\def\vkick{\ifmmode v_\mathrm{kick} \else $v_\mathrm{kick}$ \fi}
\def\vkickz{\ifmmode v_{\mathrm{kick},z} \else $v_{\mathrm{kick},z} $ \fi}
\def\vkicky{\ifmmode v_{\mathrm{kick},y} \else $v_{\mathrm{kick},y} $ \fi}
\def\vchar{\ifmmode v_\mathrm{char} \else $v_\mathrm{char}$ \fi}
\def\eflare{\ifmmode E_\mathrm{flare} \else $E_\mathrm{flare}$ \fi}
\def\ekick{\ifmmode E_\mathrm{kick} \else $E_\mathrm{kick}$ \fi}
\def\ecoll{\ifmmode E_\mathrm{coll} \else $E_\mathrm{coll}$ \fi}
\def\ezero{\ifmmode E_\mathrm{0} \else $E_\mathrm{0}$ \fi}
\def\efac{\ifmmode \xi_\mathrm{E} \else $\xi_\mathrm{E}$ \fi}
\def\tqso{\ifmmode t_\mathrm{QSO} \else $t_\mathrm{QSO}$ \fi}
\def\tflare{\ifmmode t_\mathrm{flare} \else $t_\mathrm{flare}$ \fi}
\def\tzero{\ifmmode t_\mathrm{0} \else $t_\mathrm{0}$ \fi}
\def\tfac{\ifmmode \xi_\mathrm{t} \else $\xi_\mathrm{t}$ \fi}
\def\gfac{\ifmmode f_\mathrm{g} \else $f_\mathrm{g}$ \fi}
\def\lflare{\ifmmode L_\mathrm{flare} \else $L_\mathrm{flare}$ \fi}
\def\fflare{\ifmmode F_\mathrm{flare} \else $F_\mathrm{flare}$ \fi}
\def\nflare{\ifmmode N_\mathrm{flare} \else $N_\mathrm{flare}$ \fi}
\def\tshock{\ifmmode T_\mathrm{shock} \else $T_\mathrm{shock}$ \fi}
\def\rmin{\ifmmode R_\mathrm{1} \else $R_\mathrm{1}$ \fi}
\def\rmax{\ifmmode R_\mathrm{2} \else $R_\mathrm{2}$ \fi}
\def\rbound{\ifmmode R_\mathrm{b} \else $R_\mathrm{b}$ \fi}
\def\pbound{\ifmmode P_\mathrm{b} \else $P_\mathrm{b}$ \fi}
\def\mbound{\ifmmode M_\mathrm{b} \else $M_\mathrm{b}$ \fi}
\def\mbo{\ifmmode M_{\mathrm{b}0} \else $M_{\mathrm{b}0} $ \fi}
\def\ebo{\ifmmode E_{\mathrm{b}0} \else $E_{\mathrm{b}0} $ \fi}
\def\efinal{\ifmmode E_\mathrm{final} \else $E_\mathrm{final} $ \fi}
\def\tbound{\ifmmode t_\mathrm{b} \else $t_\mathrm{b}$ \fi}
\def\tagn{\ifmmode t_\mathrm{AGN} \else $t_\mathrm{AGN}$ \fi}
\def\torb{\ifmmode t_\mathrm{orb} \else $t_\mathrm{orb}$ \fi}
\def\tdf{\ifmmode t_\mathrm{df} \else $t_\mathrm{df}$ \fi}
\def\rlim{\ifmmode R_\mathrm{lim} \else $R_\mathrm{lim}$ \fi}
\def\vlim{\ifmmode v_\mathrm{lim} \else $v_\mathrm{lim}$ \fi}
\def\vphi{\ifmmode v_\phi \else $v_\phi$ \fi}
\def\mlim{\ifmmode M_\mathrm{lim} \else $M_\mathrm{lim}$ \fi}
\def\tlim{\ifmmode t_\mathrm{lim} \else $t_\mathrm{lim}$ \fi}
\def\llim{\ifmmode L_\mathrm{lim} \else $L_\mathrm{lim}$ \fi}
\def\fqso{\ifmmode f_\mathrm{QSO} \else $f_\mathrm{QSO}$ \fi}

\def\hbeta{\ifmmode \rm{H}\beta \else H$\beta$\fi}
\def\hbetan{\ifmmode \rm{H}\beta_{\rm n} \else H$\beta_{\rm n}$\fi}
\def\hgamma{\ifmmode \rm{H}\gamma \else H$\gamma$\fi}
\def\hdelta{\ifmmode \rm{H}\delta \else H$\delta$\fi}
\def\hepsilon{\ifmmode \rm{H}\epsilon \else H$\epsilon$\fi}
\def\hzeta{\ifmmode \rm{H}\zeta \else H$\zeta$\fi}
\def\halpha{\ifmmode \rm{H}\alpha \else H$\alpha$\fi}
\def\lalpha{\ifmmode \rm{Ly}\alpha \else Ly$\alpha$}

\def\dvhb{\ifmmode \Delta v_{\hbeta} \else $\Delta v_{\hbeta}$\fi}
\def\dvmg{\ifmmode \Delta v_{\rm{Mg}} \else $\Delta v_{\rm{Mg}}$\fi}

\def\muobs{\ifmmode {\mu_{o}} \else  $\mu_{o}$ \fi}
\def\cosi{\ifmmode {\mathrm{cos}\,i} \else $\mathrm{cos}\,i$\fi}

\def\teff{\ifmmode {T_{eff}} \else $T_{eff}$ \fi}
\def\tmax{\ifmmode {T_{max}} \else $T_{max}$ \fi}

\def\tauh{\ifmmode {\tau_{\rm H}} \else $\tau_{\rm H}$ \fi}

\def\yr{\ifmmode {\rm yr} \else  yr \fi}
\def\kms{\ifmmode \rm km~s^{-1}\else $\rm km~s^{-1}$\fi}
\def\cm{\ifmmode {\rm cm} \else  cm \fi}
\def\cmmitwo{\ifmmode \rm cm^{-2} \else $\rm cm^{-2}$\fi}
\def\cmmithree{\ifmmode \rm cm^{-3} \else $\rm cm^{-3}$\fi}
\def\cmps{\ifmmode \rm cm~s^{-1}\else $\rm cm~s^{-1}$\fi}
\def\cmpsps{\ifmmode \rm cm~s^{-2}\else $\rm cm~s^{-2}$\fi}
\def\kmps{\ifmmode \rm km~s^{-1}\else $\rm km~s^{-1}$\fi}
\def\kmpspmpc{\ifmmode \rm km~s^{-1}~Mpc^{-1} \else
    $\rm km~s^{-1}~Mpc^{-1}$\fi}
  
\def\gcmthree{\ifmmode \rm g~cm^{-3} \else $\rm g~cm^{-3}$\fi}
\def\gcmtwo{\ifmmode \rm g~cm^{-2} \else $\rm g~cm^{-2}$\fi}
   
\def\erg{\ifmmode {\rm erg} \else $\rm erg$ \fi}
\def\ergps{\ifmmode {\rm erg~s^{-1}} \else $\rm erg~s^{-1}$ \fi}
\def\ergcms{\ifmmode \rm erg~cm^{-2}~s^{-1} \else $\rm erg~cm^{-2}~s^{-1}$ \fi}
\def\ergcmshz{\ifmmode \rm erg~s^{-1}~cm^{-2}~Hz^{-1} \else $\rm
erg~cm^{-2}~s^{-1}~Hz^{-1}$ \fi}
\def\ergcmsa{\ifmmode \rm erg~cm^{-2}~s^{-1}~\AA^{-1} \else $\rm
erg~cm^{-2}~s^{-1}~\AA^{-1}$ \fi}
\def\ergshz{\ifmmode \rm erg s^{-1} Hz^{-1} \else
   $\rm erg s^{-1} Hz^{-1}$ \fi}

\def\lam{\ifmmode {\lambda} \else {$\lambda$} \fi}
\def\llam{\ifmmode {L_\lambda} \else  $L_\lambda$ \fi}
\def\lamLlam{\ifmmode \lambda L_{\lambda}(5100) \else {$\lambda L_{\lambda}(5100)$} \fi}
\def\nuLnu{\ifmmode \nu L_{\nu}(5100) \else {$\nu L_{\nu}(5100)$} \fi}
\def\ilam{\ifmmode {I_\lambda} \else  $I_\lambda$ \fi}
\def\flam{\ifmmode {F_\lambda} \else  $F_\lambda$ \fi}
\def\inu{\ifmmode {I_\nu} \else  $I_\nu$ \fi}
\def\fnu{\ifmmode {F_\nu} \else  $F_\nu$ \fi}
\def\bnu{\ifmmode {B_\nu} \else  $B_\nu$ \fi}

\def\msigma{\ifmmode M_{\sigma} \else $M_{\sigma}$\fi}
\def\mbulge{\ifmmode M_{\mathrm{bulge}} \else $M_{\mathrm{bulge}}$\fi}
\def\mgal{\ifmmode M_{\mathrm{gal}} \else $M_{\mathrm{gal}}$\fi}
\def\lgal{\ifmmode L_{\mathrm{gal}} \else $L_{\mathrm{gal}}$\fi}
\def\lbulge{\ifmmode L_{\mathrm{bulge}} \else $L_{\mathrm{bulge}}$\fi}
\def\mgalstar{\ifmmode M^*_{\mathrm{gal}} \else $M^*_{\mathrm{gal}}$\fi}

\def\mbhsigstar{\ifmmode M_{\mathrm{BH}} - \sigma_* \else $M_{\mathrm{BH}} - \sigma_*$ \fi}
\def\deltalogmbh{\ifmmode \Delta~{\mathrm{log}}~M_{\mathrm{BH}} \else $\Delta$~log~$M_{\mathrm{BH}}$\fi}

\def\sigstar{\ifmmode \sigma_* \else $\sigma_*$\fi}
\def\sigthree{\ifmmode \sigma_{\mathrm{[O~III]}} \else $\sigma_{\mathrm{[O~III]}}$\fi}
\def\sigtwo{\ifmmode \sigma_{\mathrm{[O~II]}} \else $\sigma_{\mathrm{[O~II]}}$\fi}
\def\signl{\ifmmode \sigma_{\mathrm{NL}} \else $\sigma_{\mathrm{NL}}$\fi}
\def\wthree{\ifmmode {\rm FWHM({[O~III]})} \else $FWHM({[O~III]})$ \fi}
\def\wtwo{\ifmmode {\rm FWHM({[O~II]})} \else $FWHM({[O~II]})$ \fi}
\def\mthree{\ifmmode M_{\mathrm [O~III]} \else $M_{\mathrm [O~III]}$ \fi}
\def\mtwo{\ifmmode M_{\mathrm [O II]} \else $M_{\mathrm [O II]}$ \fi}

\def\lbreak{\ifmmode L_{\mathrm{break}} \else $L_{\mathrm{break}}$\fi}
\def\lcut{\ifmmode L_{\mathrm{cut}} \else $L_{\mathrm{cut}}$\fi}

\received{}
\accepted{}

\slugcomment{Submitted to ApJ}

\shortauthors{Smith, Shields, Bonning, McMullen, Rosario \& Salviander}
\shorttitle{Binary QSOs}

\begin{document}

\title{LBQS 0103-2753: A Binary Quasar in a Major Merger}

\author{G.~A. Shields\altaffilmark{1}, D.~J. Rosario\altaffilmark{2}, V. Junkkarinen\altaffilmark{3}, S.~C. Chapman\altaffilmark{4}, 
    E.~W. Bonning\altaffilmark{5},  T. Chiba\altaffilmark{6} }

\altaffiltext{1}{Department of Astronomy, University of Texas, Austin,
TX 78712; shields@astro.as.utexas.edu} 

\altaffiltext{2}{Max Planck Institute for Extraterrestrial Physics, Garching, 85748, Germany; rosario@mpe.mpg.de}

\altaffiltext{3}{CASS, University of California, San Diego, La Jolla, CA 92093; vesa@ucsd.edu}

\altaffiltext{4}{Institute of Astronomy, University of Cambridge, Madingley Road, Cambridge CB3 0HA, U.K.; schapman@ast.cam.ac.uk}

\altaffiltext{5}{YCAA - Department of Physics, Yale University, New Haven, CT 06520; erin.bonning@yale.edu}

\altaffiltext{6}{Department of Applied Physics, Yale University, New Haven, CT 06520; tamara.chiba@yale.edu}

\begin{abstract}

We present \hst\ and UKIRT spectra and images of the 2~kpc binary quasar LBQS~0103-2753  ($z=0.858$).  The \hst\ images
(V- and I-band) show tidal features demonstrating that this system is a major galaxy merger in progress.  A two-color composite image brings out knots of star formation along the tidal arc and elsewhere.  The infrared spectrum shows that both objects are at the same redshift, and that the discrepant redshift of \civ\ in component A is a consequence of the BAL absorption in the spectrum of this component.

LBQS 0103-2753 is one of the most closely spaced binary QSOs known, and is one of relatively few dual AGN showing confirmed broad emission lines from both components.  While statistical studies of binary QSOs suggest  that simultaneous fueling of both black holes during a merger may be relatively rare, LBQS 0103-2753 demonstrates that such fueling can occur at high luminosity at a late stage in the merger at nuclear spacing of only a few kpc, without severe obscuration of the nuclei.

\end{abstract}


\section{Introduction}
\label{sec:intro}

Dual AGN, representing merging galaxies with both nuclei in an active state, are a subject of increasing interest.  Dual AGN at spacings of 10 to 100~kpc, spatially resolved from the ground, have been thought to occur with a frequency of about 1 per thousand AGN \citep{kochanek99, hennawi06}.   However, a recent study by
\citet{liu11} finds a frequency as much as 10 times higher than this for AGN in the Sloan Digital Sky Survey (SDSS)\footnote{The SDSS website is http://www.sdss.org.}.
Several more closely spaced dual AGN are known from a variety of observational techniques.   \hst\ imaging has revealed  LBQS~0103-2753  at 2~ kpc \citep{junkkarinen01} and \cosmos\  at a similar spacing \citep{comerford09, civano11} .  A close pair observed
in X-rays is the ULIRG NGC 6240 at a spacing of 1.4~kpc \citep{komossa03}.  The closest known binary black hole, spaced by 7~pc, is in the radio galaxy 0402+379  \citep{rodriguez09}.   \citet{junkkarinen01}estimated a frequency of 1 dual QSO per 1000 QSOs at spacings $\sim2$~kpc.  They argued that this was surprisingly low if one assumes that most luminous AGN are triggered by 
mergers and that fueling of one black hole is often accompanied by fueling of the other.   A similar  conclusion was reached for spatially resolved dual AGN in a theoretical study by \citet{foreman09}.
In the last several years, a number of studies of dual AGN have appeared, including systematic searches for candidate objects based
on spectroscopic indicators \citep{wang09, liu10a, smith10} and follow-up imaging and spectroscopic studies \citep{liu10b, fu11, rosario11}.  A detailed study of a z=0.44 dual quasar with a spacing of 21 kpc is given by \citep{green10}, and X-ray confirmation of a dual AGN at 2 kpc spacing was recently reported by \citet{comerford11}.  A majority of these objects show only narrow emission lines.  Known dual AGN showing broad emission lines from both components, as does \lbqs, are relatively rare.

The binary nature of LBQS~0103-2753 was discovered by \citet{junkkarinen01} in the course of ultraviolet spectroscopic observations with STIS aimed at understanding the broad absorption lines.  The acquisition image revealed two AGN separated by 0.3 arcsec with magnitude V = 18.2 and 19.4 for components A and B, respectively.   However, the acquisition image was too shallow to show the host galaxies.  Moreover, there was an apparent discrepancy between the redshifts of the two components.  Junkkarinen et al. assumed that this resulted from the presence of BAL features in the brighter component, which can affect the redshift of the \civ\ emission line used for the redshift.  In this paper, we present follow-up imaging and spectroscopic observations that confirm this exceptional
object as a true dual AGN in an on-going galactic merger.

 \section{Observations}
 \label{sec:obs}

 We have obtained imaging of \lbqs\ aimed at revealing the nature of the host galaxy and spectroscopy aimed at clarifying the
 redshift agreement between components A and B and testing for variability of the BAL features of component A.
 
 \subsection{\hst\ Images}
 \label{sec:hst}

LBQS 0107-2753 was imaged with the Advanced Camera for Surveys (ACS) on \hst\ using the F606W and F814W filters for the purpose of
studying the host galaxy morphology and any merger signatures (GO 9498, PI Junkkarinen).  These filters correspond roughly to rest wavelengths of 3300 and 4400~\AA, spanning the break at \caii\ H and K.   Each image is composed of 4 dithered subexposures, each of 1020s, yielding a total integration time of 4080s per filter. A 4-point box dither was used, with a box size of 0.265 arcsec. The images were combined using the standard version of MULTIDRIZZLE run by the archive pipeline. Remaining cosmic rays were removed using the CRREJECT task in IRAF\footnote{IRAF is distributed by the National Optical Astronomy Observatories, which are operated by the Association of Universities for Research in Astronomy, Inc., under cooperative agreement with the National Science Foundation.}. A conservative approach was employed in the cosmic-ray rejection algorithm so as not to remove flux from the bright QSO point source cores. Therefore, some weak CR residuals still remain in the final image.

\begin{figure}[ht]
\begin{center}
\plotone{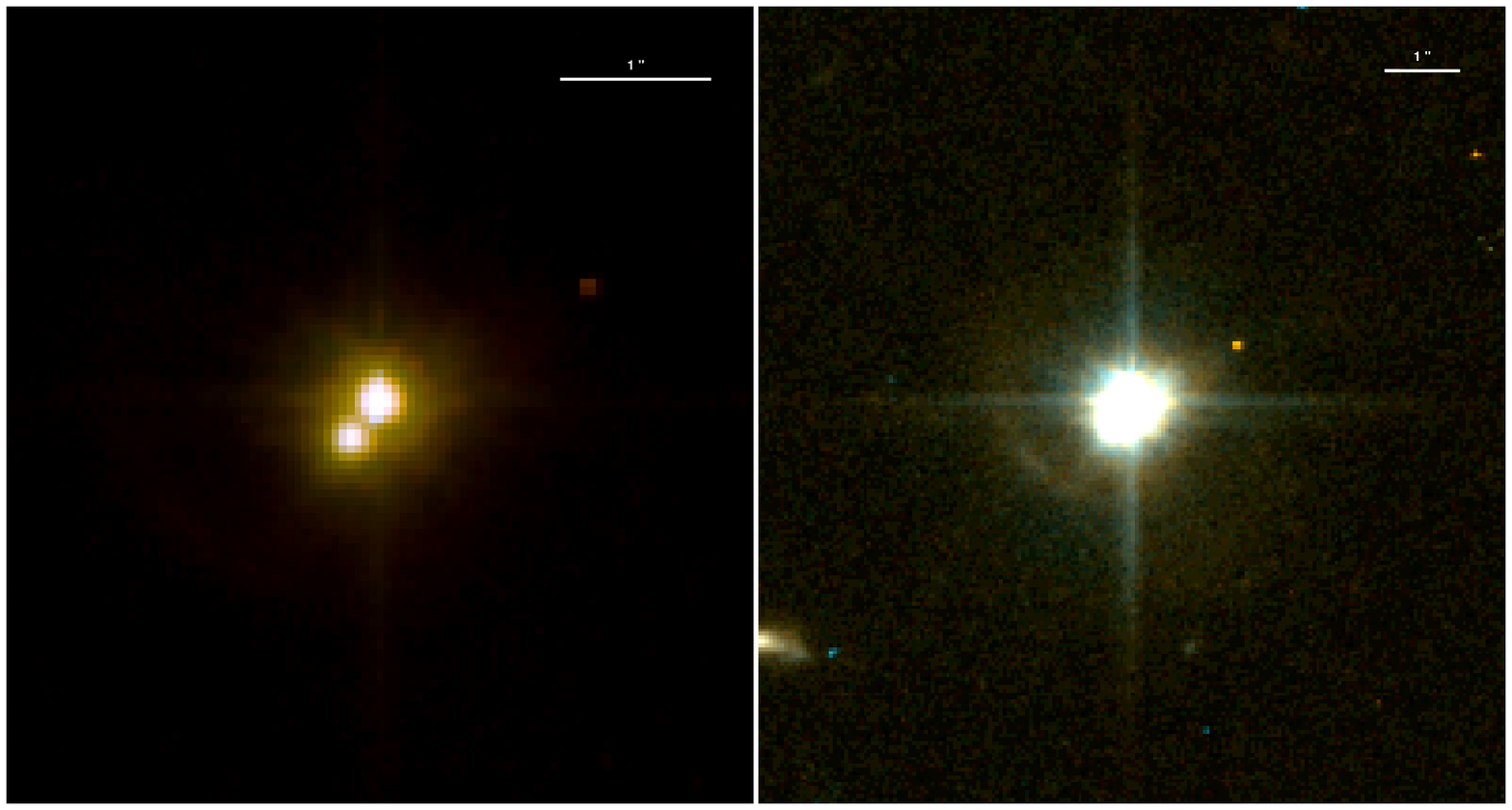}
\figcaption[]{
 Two-color image of LBQS~0103-2753 representing a combination of F606W and F814W exposures with the ACS on \hst.
 The left panel gives a shallower rendition at a larger spatial scale to display the two QSOs.  The right panel gives
 a deeper image to bring out the host galaxy and tidal arc.
See text for discussion.
\label{fig:image} }
\end{center}
\end{figure}

Figure \ref{fig:image} shows a two color rendition of the HST images.  The left panel brings out the two AGN.  The right panel is zoomed out a factor of two and gives a deeper rendition,  clearly showing the host galaxy.  There is a prominent tidal arc at about 10~kpc from the nucleus that can be traced from position angles roughly 30\deg\ to 180\deg, reaching a distance of $\sim15$~kpc at the northeast extreme.  A possible companion galaxy is visible 6~arcsec SE of the quasar, but the full image does not indicate the presence of
a rich cluster.

\begin{figure}[ht]
\begin{center}
\plotone{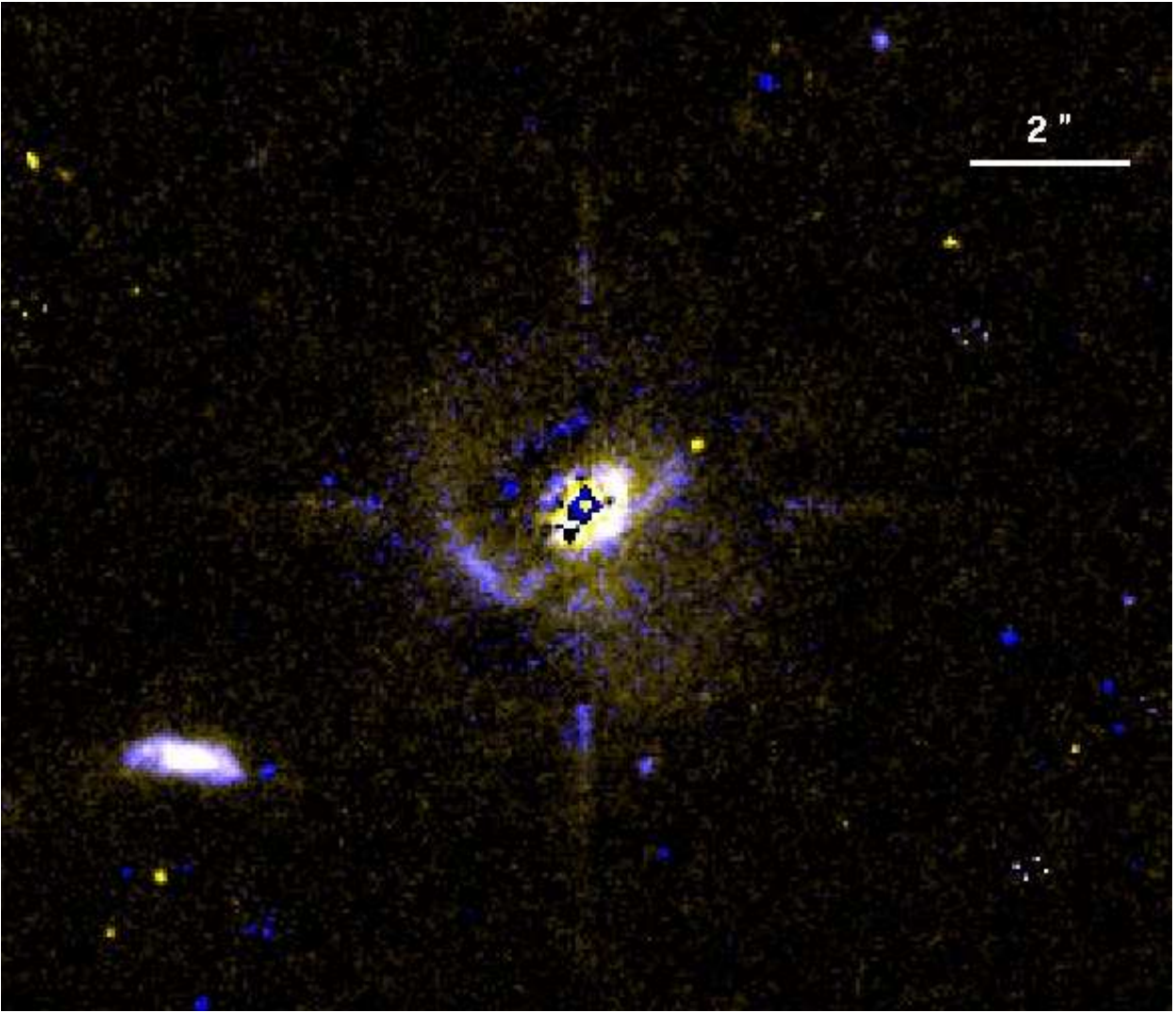}
\figcaption[]{
PSF-subtracted two-color image of LBQS~0103-2753 weighted to bring out the regions of recent star formation
in the main tidal arc and the counter-arc.
See text for discussion.
\label{fig:qsosub} }
\end{center}
\end{figure}

In an effort to bring out the host galaxy morphology, the light of the two quasars was subtracted using a PSF derived from neighboring stars in each image.  Figure \ref{fig:qsosub} shows a difference image with F606W in blue and F814W in yellow, made with the PSF-subtracted images.  This brings out several regions of intense star formation that are strong in the shorter wavelength exposure, including the main tidal arc, a smaller counter-arc, and some other scattered patches with a bluer stellar population.   The tidal features and disturbed morphology confirm that the system is a major merger in progress.  The tidal arm structure is somewhat reminiscent of NGC4038/39 (the ``Antennae''); however, that system has a nuclear spacing of $\sim 6$~kpc \citep{stanford90}, about 3 times larger than for \lbqs.  

There is a diffuse envelope of starlight visible to radii $\sim20$ to 25~kpc from the nucleus.   In order to study the light profile of this envelope, we performed annular photometry on the I-band image in the radial range from 0\farcs5 to 4\farcs0.  The profile is well fitted by a S\'ersic function with an index of 0.6 and half-light radius of 1\farcs8 (14~kpc).  This very flat profile is suggestive of a disk but might also represent a distended envelope resulting from the merger.    An integration of the best fit profile over all radii gives a total flux in F814W, which translates to an absolute blue magnitude of $M_B = -23.0$.  This is about 10\%\ of the total light from the QSOs.  If this light is largely contributed by bulge stars from the progenitor galaxies, it is roughly consistent with the sum of the black hole masses for the QSOs estimated below \citep{kormendy01}.  We therefore have a picture in which two rather large galaxies with commensurate black holes are being observed at an advanced stage of their merger.   The exceptional nature of the present system lies in the vigorous, simultaneous fueling of both black holes, and the lack of obscuration of either nucleus, at least on our line of sight.

 \subsection{Infrared Image}
 \label{sec:irimage}

In the course of acquiring infrared spectra with UKIRT (see below), we obtained a K-band image in approximately 0.5~arcsec seeing.
The image is elongated in the position angle of the binary AGN, showing that the two nuclei are marginally resolved.
Modeling the image with two point sources based on the PSF of nearby stars gives magnitudes of K = 16.46 and 17.85 for
components A and B, respectively, with an uncertainty of about 0.1~magn.  Absent brightness variations since the Junkkarinen et al (2001) observations, this gives colors of V - K = -1.7 and -1.6  for A and B, respectively, so that there is little difference in optical - infrared color between the two components.  

 \subsection{Infrared Spectrum}
 \label{sec:irspec}
 
\citet{junkkarinen01} found a discrepancy of $\sim4000~\kms$ between the redshifts of components A and B, based on the peak of the \civ ~1549~\AA\ emission line ($z_A = 0.834$ and $z_B = 0.848$).  They noted, however, that the \civ\ emission line peak can be substantially skewed in BAL QSOs.   The \halpha\ emission line should not be affected by the BALs.   In order to test for the redshift agreement of components A and B, we obtained a J-band infrared spectrum with the United Kingdom Infrared Telescope (UKIRT) on the night of 2001 August 28.  The setup employed the CGS4 spectrograph with a grating of 40 lines per mm and a 2 pixel slit (1.2 arcsec), giving resolution R/400.  The detector was a 256 by 256 array.  A total of 12 exposures of 60~sec were obtained using alternate chops along the slit.  Standard calibration procedures were carried out using the XIDL package (http://www.ucolick.org/~xavier/IDL).
 
\begin{figure}[ht]
\begin{center}
\plotone{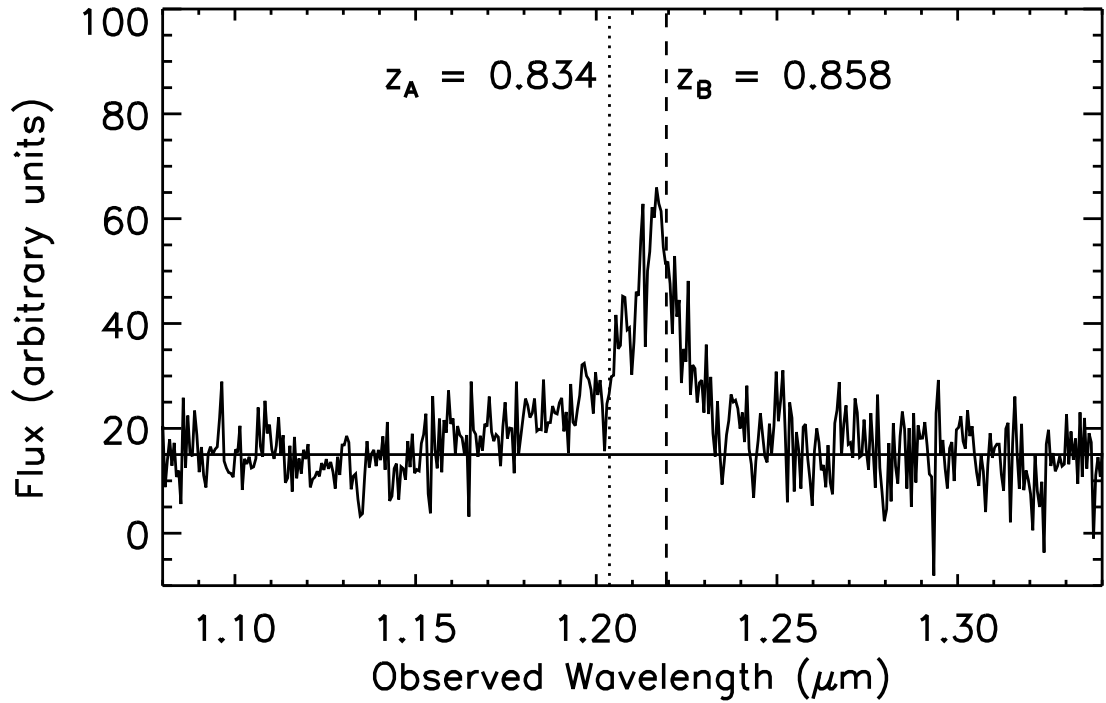}
\figcaption[]{
Infrared J-band spectrum of LBQS~0103-2753 obtained with UKIRT.  The expected wavelengths of \halpha\ for the 
redshift of the \civ\ peak in  component A ($z=0.834$) and of component B ($z = 0.858$) are shown.  There is no indication
of a feature at $z_{\mathrm A}$, confirming that the \civ\ peak in component A  is affected by the BAL absorption and does not
represent the true redshift.
See text for discussion.
\label{fig:irspec} }
\end{center}
\end{figure}

 The reduced J-band spectrum is shown in Figure \ref{fig:irspec}.  This spectrum includes the light from both components of the dual QSO.  The broad \halpha\ line is evident.  The profile suggests a single redshift, rather than a superposition of two lines at $z_A$ and $z_B$.  The wavelength of the \halpha\ peak is  12,175~\AA, giving a redshift of $z = 0.855$.  This is in good agreement
 with $z_B = 0.858$.  The expected wavelength of \halpha\ for redshift $z_A$ is 12,036~A, which is distinctly displaced to the blue wing of the
 observed profile.  Since component A is the brighter component, one would expect the combined \halpha\ profile to be dominated by
 component A.  We conclude that the \civ\ redshift of component A is indeed affected by the BAL, and that
 the true redshift of component A agrees with component B.  In view of the low probability of a chance superposition \citep{junkkarinen01} and the tidal features evident in the images, we conclude that this system is indeed a dual AGN in an ongoing merger.

 \subsection{\hst\ Spectra}
 \label{sec:hstspec}
 
\begin{figure}[ht]
\begin{center}
\plotone{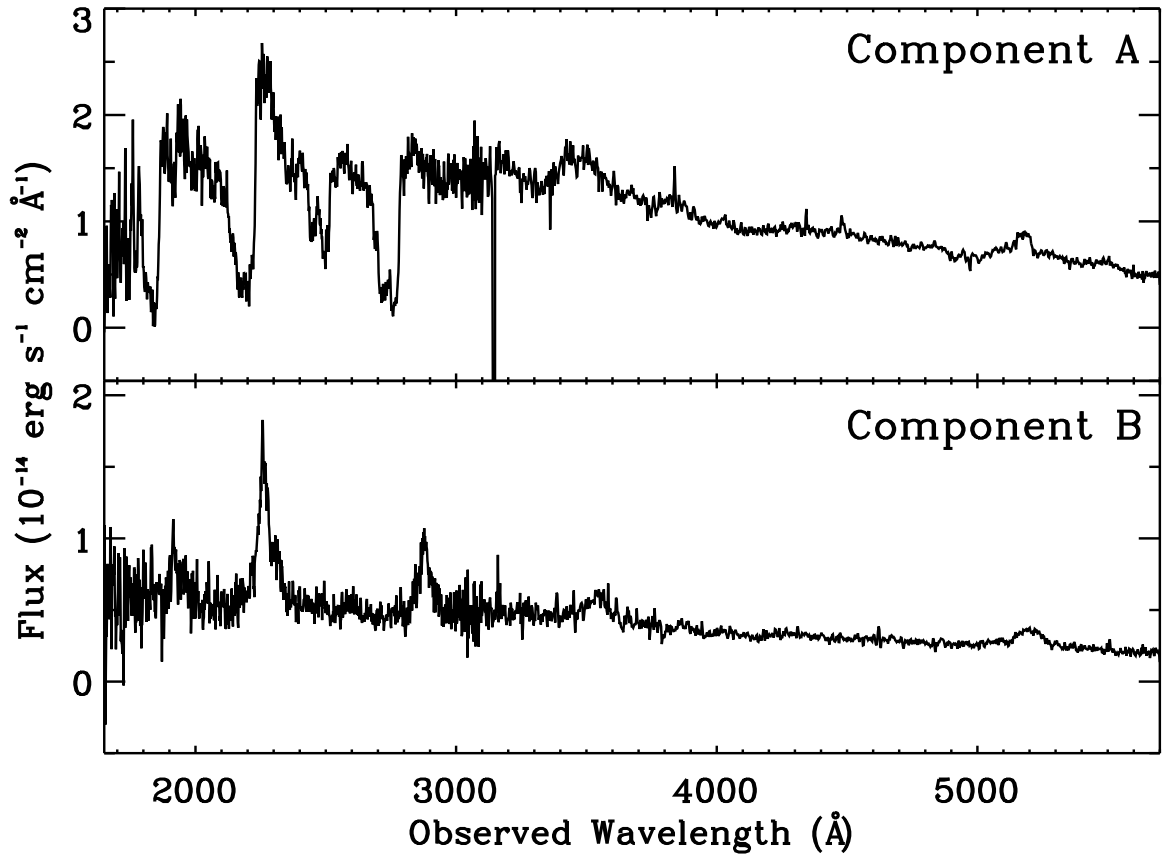}
\figcaption[]{
\hst\ STIS spectra of LBQS~0103-2753 for components A and B.  See text for discussion.
\label{fig:hstspec} }
\end{center}
\end{figure}

Spectra of \lbqs\ in the ultraviolet and optical were obtained with the Space Telescope Imaging Spectrograph (STIS) on July 18 and 20, 2002. Three gratings were used:  G230L (0\farcs2 slit width, exposure 2359 s), G430L( 0\farcs1, 2856 s), and G750L (0\farcs1, two exposures of 2672~s and 1834~s).  The slit was oriented at position angle 150\deg\ to capture both components of the QSO.   Figure \ref{fig:hstspec} shows the flux-calibrated spectra from G230L and G430L from the \hst\ pipeline combined into a single spectrum joined at 3100~\AA.   The spectrum shows the prominent BAL features associated with the \ovi, \lalpha, Si~IV, and \civ\ permitted lines.  Comparison with Figure 2 of \citet{junkkarinen01} shows little variation in the emission lines or BAL features.  The G430L spectrum includes the \ciii\ and \mgii\ lines.   The \mgii\ BAL remains weak, similar to the case for 1998 CTIO spectrum obtained by \citet{junkkarinen01}.  This contrasts with the strong \mgii\ BAL in the LBQS spectrum from 1987 \citep{morris91}.

\begin{figure}[ht]
\begin{center}
\plotone{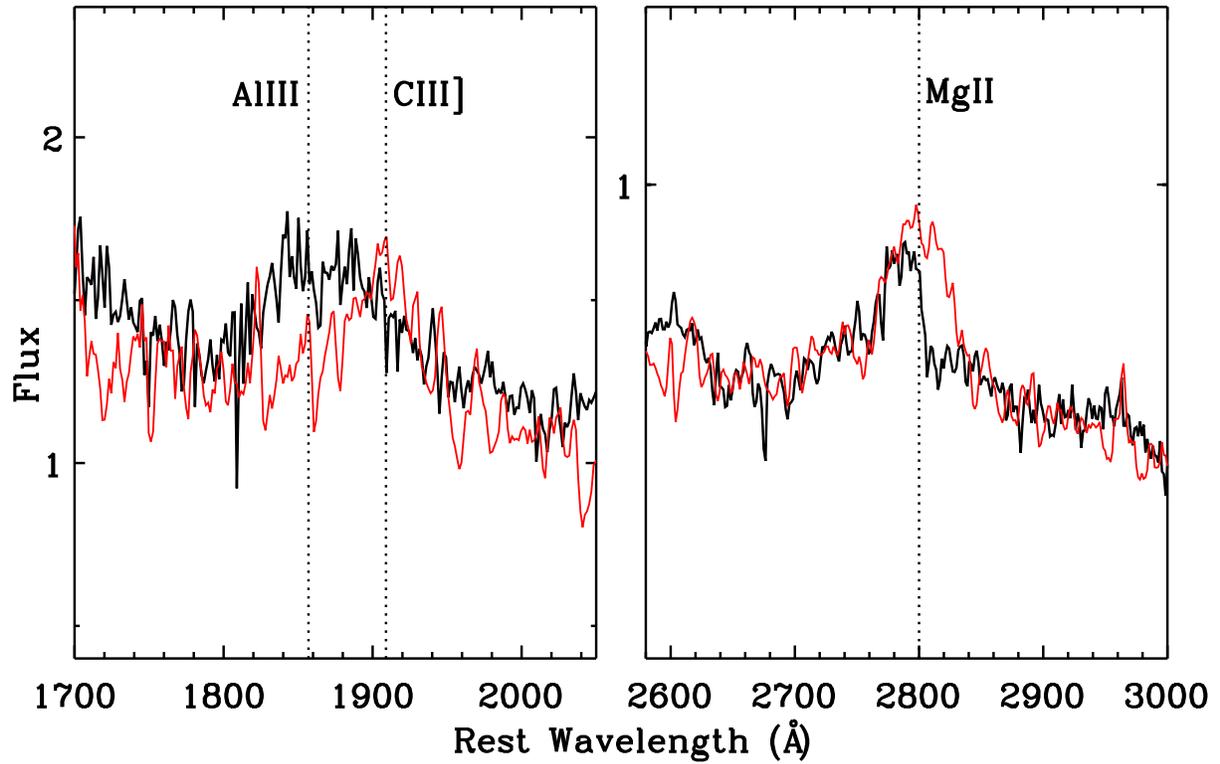}
\figcaption[]{
Overlay of the STIS spectra of LBQS~0103-2753 for components A (black) and B (red).  Fluxes have been scaled to overlap in the continuum.  Vertical dotted lines show expected wavelengths for $z_B = 0.858$.  Note the wavelength offset of \mgii\ in component A and the strong emission in Al~III $\lambda1857$.  See text for discussion.
\label{fig:alum} }
\end{center}
\end{figure}

Figure \ref{fig:alum} shows a comparison of the spectra of components A and B in the \ciii\ and \mgii\ regions.  Component A shows strong emission in Si~III~$\lambda1892$ and  Al~III~$\lambda1857$.  Strong Al~III emission is common in BAL QSOs \citep{hartig86}.  There are also substantial differences between the emission profiles of \mgii\ in the two QSOs, including a blueward shift of $\sim1500~\kms$ in the \mgii\ peak in component A relative to component B.

The goal of the G750L spectra was to measure the \hbeta\ profile and detect any narrow \oiii\ lines that would elucidate the redshift of components A and B.  Unfortunately, no \oiii\ lines were detected.  The broad \hbeta\ lines of both components are  detected but too noisy for useful measurements.

 \section{Discussion}
  \label{sec:discuss}
 
Our observations confirm that LBQS~0103-2753 is indeed a dual quasar in an on-going major merger.  This object is distinguished from
other known examples by the combination of close spacing, substantial redshift, and the high luminosity.   Other examples of dual AGN
with good imaging of the host galaxy include \cosmos\ and  the z=0.44 dual AGN  SDSS J1254+0846 separated by 21 kpc \citep{green10}.  
The latter object is more widely spaced than our object, and the tidal tail is more extensive ($\sim75$~kpc)  and symmetrical.

From a suite of numerical simulations of mergers, \citet{green10}) find a good fit to the observed geometry of SDSS J1254+0846 for a prograde merger on its second pass with a 2-to-1 baryon mass ratio.   From the width of the  broad \hbeta\ line of each component of SDSS J1254+0846, Green et al. derive a black hole mass $\mbh = 10^{8.46}, 10^{7.80}~\msun$ for components A, B.  This gives Eddington ratios of 0.67 and 0.29 respectively.  The former value and marginally the latter one are higher than typical for AGN of this luminosity and redshift, which as Green et al. suggest may reflect vigorous fueling in the on-going merger.   Green et al. infer from the merger simulations that the incoming galaxies must have had large disks to produce the large tidal tails and had pre-existing bulges to give the large measured black hole masses at the early merger stage of SDSS J1254+0846.  For comparison, \citet{junkkarinen01}  estimated $\mbh = 10^{9.0}, 10^{8.5}~\msun$ for components A, B in \lbqs\ on the assumption that each component is shining at one-third of the Eddington limit.  Alternatively, for component B, the broad line widths may be used to estimate \mbh.  From the \hst\ spectrum, the FWHM of \civ\ is $\sim5200~\kms$ and the luminosity at 1350~\AA\ is $\nuLnu \approx 10^{45.18}~\ergps$.  Using the expression for \mbh\ in \citet{vestergaard06}, we find $\mbh \approx  10^{8.7}~\msun$, giving $L/\led = 10^{-0.7}$.  By either approach, \lbqs\ involves luminosities and black hole masses one-half to one order of magnitude larger than for SDSS J1254+0846.  (\mgii\ has a FWHM of $\sim5700~\kms$, similar to \civ, but an unusually flat topped profile. Therefore we do not use it for \mbh.)

\citet{comerford09} give strong evidence for a dual AGN at $z = 0.36$ in \cosmos.   The \hst\ image shows a tidal tail and two point
bright point sources separated by 1.75~kpc, whose spectra are indicative of AGN.  The spacing of the two AGN and the asymmetrical tidal
tail of this object resemble \lbqs\ more closely than does SDSS J1254+0846.   \citet{civano11} present a multiwavelength analysis of 
this object, giving an optical luminosity for the brighter component of $\nuLnu = 10^{43.9}, 10^{43.6}~\ergps$ for the two AGN components,
about two orders-of-magnitude less luminous than \lbqs.  
For the SE (fainter) component, they estimate a black hole mass of $10^{7.8}~\msun$ from the width of the broad \hbeta\ line.   
Civano et al. make the interesting suggestion that the SE component may actually be a runaway black hole ejected from
the host galaxy nucleus by gravitational radiation recoil or the three-body slingshot mechanism.  This interpretation is based in part
on a large velocity offset ($\sim1200~\kms$) between the broad \hbeta\ line of the SE component and the narrow emission lines.

Similar in redshift and spacing to \lbqs\ is the dual AGN \citep{gerke07}.  This object appears to be in an early type host with some
morphological disturbance.  While we cannot analyze the morphology of the host of \lbqs\ in detail, the large tidal arc and the prominent star formation regions suggest a gas-rich merger, consistent with the vigorous fueling required for  the two luminous QSOs.

\acknowledgments

We thank Karl Gebhardt for helpful discussions and assistance.  G.S. gratefully acknowledges support from the Jane and Roland Blumberg Centennial Professorship in Astronomy at the University of Texas at Austin.

The United Kingdom Infrared telescopes is operated by the Joint Astronomy Centre on behalf of the science and Technology Facilities of the U.K.  Funding for the Sloan Digital Sky Survey (SDSS) has been provided by the Alfred P. Sloan Foundation, the Participating Institutions, the National Aeronautics and Space Administration, the National Science Foundation, the U.S. Department of Energy, the Japanese Monbukagakusho, and the Max Planck Society. The SDSS Web site is http://www.sdss.org/. The SDSS is managed by the Astrophysical Research Consortium (ARC) for the Participating Institutions. The Participating Institutions are The University of Chicago, Fermilab, the Institute for Advanced Study, the Japan Participation Group, The Johns Hopkins University, the Korean Scientist Group, Los Alamos National Laboratory, the Max-Planck-Institute for Astronomy (MPIA), the Max-Planck-Institute for Astrophysics (MPA), New Mexico State University, University of Pittsburgh, University of Portsmouth, Princeton University, the United States Naval Observatory, and the University of Washington.

\clearpage


\newpage


\newpage


\newpage


\newpage


\newpage


\begin{thebibliography}{38}

\expandafter\ifx\csname natexlab\endcsname\relax\def\natexlab#1{#1}\fi

\bibitem[Civano et al.(2011)]{civano11} Civano, F., \etal\ 2011, \apj, 717, 209

\bibitem[Comerford et al.(2009)]{comerford09} Comerford, J.M. \etal\ 2009, \apjl, 702, L82

\bibitem[Comerford et al.(2011)]{comerford11} Comerford, J.M., Pooley, D., Gerke, B. F., \& Madejski, G. M. 2011, \apjl, 737, L19

\bibitem[Foreman et al.(2009)]{foreman09} Foreman, G., Volonteri, M., \& Dotti, M.  2009, \apj, 693, 1554

\bibitem[Fu et al.(2011)]{fu11} Fu, H., et al. 2011, preprint (arXiv:1107.3564)

\bibitem[Gerke et al.(2007)]{gerke07} Gerke, B. F. \etal\ 2000, \apjl, 660, L23

\bibitem[Green et al.(2010)]{green10} Green, P., et al. 2010, \apj, 712, 762

\bibitem[Hartig \& Baldwin(1986)]{hartig86} Hartig, G. F., \& Baldwin, J. A. 1986, \apj, 302, 64

\bibitem[Hennawi(2006)]{hennawi06} Hennawi, J. F., et al. 2006, \aj, 131, 1

\bibitem[Junkkarinen et al.(2001)]{junkkarinen01} Junkkarinen, V.  \etal\ 2001, \apj, 549, L155

\bibitem[Kochanek(1999)]{kochanek99} Kochanek, C. S. , Falco, E. E., \& Mu\~oz, J. A. 1999, \apj, 510, 590

\bibitem[Komossa(2003)]{komossa03} Komossa, S., et al.  2003,  \apjl, 582, L15

\bibitem[Kormendy \& Gebhardt (2001)]{kormendy01}
{Kormendy}, J. \& {Gebhardt}, K. 2001, in AIP Conf. Proc. 586: 20th Texas  Symposium on relativistic astrophysics, ed. H.~Martel \& J.~C. Wheeler,  363

\bibitem[Liu et al.(2010a)]{liu10a} Liu, X., Shen, Y., Strauss, M. A., \& Greene, J. E. 2010, \apj, 708, L427

\bibitem[Liu et al.(2010b)]{liu10b} Liu, X., Greene, J. E., Shen, Y., Strauss, M. A., \&  2010, \apj, 715, L30

\bibitem[Liu et al.(2011)]{liu11} Liu, X., Shen, Y., Strauss, M. A., \& Hao, L. 2011, \apjl, 737, L101

\bibitem[Morris et al.(1991)]{morris91} Morris, . S. L., et al. 1991, \aj, 102, 1627

\bibitem[Rodriguez et~al.(2009)] {rodriguez09} Rodriguez, C., Taylor, G. B., Zavala, R. T., 
  Philstro\"om, Y. M., \& Peck, A. N. 2009, apj, 697, 37

\bibitem[Rosario et al.(2011)] {rosario11} Rosario, D., McGurk, R. C., Max, C. E., Shields, G. A., \& Smith, K. L. 2011, in press (arXiv:1102.1733)

\bibitem[Stanford et al.(1990)]{stanford90} Stanford, S. A., Sargent, A. I., Sanders, D. B., \& Scoville, N. Z. 1990, \apj, 349, 492.

\bibitem[Smith et al.(2010)]{smith10} Smith, K.L., Shields, G.A., Bonning, E.W., McMullen, C.C., Rosario, D.J., \& Salviander, S. 2010, \apj, 716, 866

\bibitem[Vestergaard \& Peterson(2006)]{vestergaard06}  Vestergaard, M., \& Peterson, B. M. 2006, \apj, 641, 689.

\bibitem[Wang et al.(2009)]{wang09} Wang, J., Chen, Y., Hu, C., Mao, W., Zhang, S., Bian, W. 2009, \apjl, 705, L76

\end{thebibliography}
\end{document}